# Full-vectorial mode coupling in optical fibers

Liang Fang, Jian Wang*, *Member, IEEE*

*Abstract*—Optical full-vectorial modes have closely attracted much attention because of their unique optical features and thus may develop the fundamental researches and new functionalities of light. Beyond the conventional linear polarization (LP) modal coupling, we thoroughly investigate the inherent polarization-dependence and polarization-isolation features for full-vectorial mode coupling in optical fibers, as well as the coupling crosstalk in view of the modal degeneracy. We present the general full-vectorial coupling mechanism and numerically simulate the full-vectorial mode coupling for both fiber directional and grating-assisted coupling cases. Furthermore, we give the detailed analyses of coupling crosstalk based on the unified three-mode coupling model. Apart from lifting the modal degeneracy, significantly, the crosstalk can also be reduced by weakening the coupling coefficient and meanwhile increasing the coupling length. This extended full-vectorial mode coupling mechanism may provide theoretical guidance on exploiting new all-fiber-based polarization-sensitive passive elements to control full-vectorial fields and modal polarization states.

## I. INTRODUCTION

The space-varying full-vectorial dimensionality of optical wave has been increasingly studied in recent years [1-3], since the conventional optical degrees of freedom were full exploited, including amplitude and wavelength, uniform phase and polarization. As an additional available degree of freedom, the full-vectorial fields characterized by spatially inhomogeneous electric-field vector may develop the fundamental researches and new functionalities of light, and potentially apply to particle acceleration [4], optical trapping and microscopy [5,6], optical communications and sensing [7,8], and etc. Apart from existing in free space, when confined by optical waveguides with boundary constraint, the electromagnetic-vector fields could also exhibit spatial inhomogeneous distributions, especially in the optical fibers because of the circular symmetry of waveguide structures [9]. Generally, the vectorial fields with higher-order states would degenerate into linear polarization (LP) multi-lobes modes in weakly guiding conditions. However, in high-contrast-index optical fibers [10,11], the full-vectorial modes could be separated in terms of effective refractive index, for instance, the first modal group including $TM_{01}$, $TE_{01}$, and $HE_{21}$ modes.

It is indispensable to flexibly manage these exploitable full-vectorial modes, such as generation, conversion, and polarization control to advance the development of silicon photonic integrated circuits, fiber optics and their related applications. In free space, arbitrary full-vectorial fields can be obtained by interference with two orthogonal vortex fields [12], or through glass cone [2]. In optical waveguides, mode coupling is the ubiquitous way to manage the fiber-guided modes. Almost all previous reports on mode coupling focused on the LP mode in the weakly guiding fibers [13-17]. However, so far, there have been few reports on the comprehensive analyses of mode coupling between two full-vectorial modes in the strongly guiding fibers with high-contrast-index distribution, let alone the coupling crosstalk. As is well known, the polarization states of two coupled modes remain uniform for the LP mode coupling in the weakly guiding single-mode fibers (SMFs) or few-mode fibers (FMFs). Apparently, as for the spatially inhomogeneous electric-field distribution of full-vectorial modes, especially with different higher-order states, the conventional LP mode coupling principle about polarization orientation can not give a straightforward guidance on the full-vectorial mode coupling.

In this article, we study the full-vectorial coupling principle in optical fibers for both fiber directional and grating-assisted coupling cases, and investigate its inherent polarization-dependence and polarization-isolation features based on the coupled-mode theory. We numerically simulate these two full-vectorial coupling cases and analyze the coupling crosstalk in view of modal degeneracy degree of full-vectorial modes in optical fibers. By taking advantage of these coupling features, it is hopeful to achieve a new manageable degree of freedom for fiber-guided modal polarization states, compared to the previous management of optical power, wavelengths and modal forms [18-20], and thus design novel all-fiber-based polarization-sensitive passive elements to manage the available full-vectorial fields and modal polarization components.

## II. COUPLING-MODE THEORY

Coupled-mode theory is an available way to research mode evolution between different propagating modes along optical fibers caused by permittivity perturbation. As for higher-order LP mode coupling in the weakly guiding fibers [17, 22], the spatial orientation of mode intensity patterns (i.e., even or odd LP mode) depends upon the core structural layout or tilted grating orientation, without considering their uniform polarization orientation. However, the full-vectorial modes do

This paragraph of the first footnote will contain the date on which you submitted your paper for review. It will also contain support information, including sponsor and financial support acknowledgment. For example, "This work was supported in part by the U.S. Department of Commerce under Grant BS123456".

The authors are with the Wuhan National Laboratory for Optoelectronics, School of Optical and Electronic Information, Huazhong University of Science and Technology, Wuhan 430074, Hubei, China. (email: jwang@hust.edu.cn).

not manifest spatial-dependent mode intensity patterns, but are intrinsically characterized by the spatial-dependent polarization distribution and no field intensity in optical singularity. Accordingly, it may give rise to some unique coupling features for full-vectorial mode coupling in optical fibers, which is worth being comprehensively restudied, especially for the coupling crosstalk when considering the sensitive mode degeneracy from two close full-vectorial modes into LP mode.

*A. Coupling coefficient*

The space-varying electric-field vector orientation must be fully considered with the transverse refractive index perturbation for the full-vectorial mode coupling. It can be generally described by the well-known coupling coefficient, written as,

$$K = \frac{1}{4}\omega \iint_\infty \Delta\varepsilon \cdot \vec{e}_1 \cdot \vec{e}_2^* dxdy, \quad (1)$$

where $\omega$ are the angular frequency of light, $\vec{e}_1$ and $\vec{e}_2$ describe the expressions of spatially inhomogeneous transverse electric-field vectors for two coupled modes, which can be generally divided into two orthogonal vector components in the cartesian coordinates, i.e. $\vec{e}_v = \sum_q E_q^v(x,y)\vec{e}_q$ with $E_q^v$ being the transverse electric-field magnitude of each component $(v=1,2; q=x,y)$ that satisfies the Helmholtz equation [21]. If being longitudinal-independent, $\Delta\varepsilon(x,y)$ denotes the transverse refractive index perturbation to the permittivity and is proportional to the refractive index difference $\Delta n(x,y)$. Hence, the coupling coefficient becomes

$$\kappa \propto \iint_S \Delta n(x,y) \cdot \left( E_x^1 \cdot E_x^{2*} + E_y^1 \cdot E_y^{2*} \right) dxdy, \quad (2)$$

where $S$ represents the perturbation region. It should be emphasized that two modes produce coupling on the condition that their electric-field vector orientation makes the coupling coefficient maximum, which distinctly determines the polarization-dependent feature for the full-vectorial coupling.

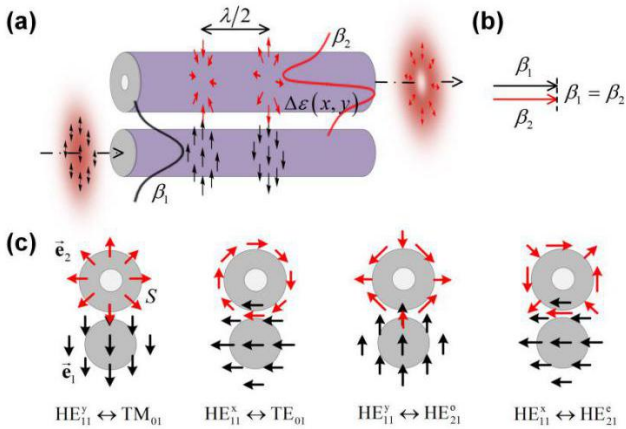

Fig. 1. The sketch and illustration of full-vectorial directional coupling. (a) Full-vectorial mode coupling from the fundamental mode $HE_{11}$ in SMF to the radially polarized mode $TM_{01}$ in RCF. (b) The resonance condition of directional coupling. (c) The relationship between electric-field vector orientation of two coupled full-vectorial modes and transverse core layout.

In general, there are two large families of fiber-guided mode coupling cases, i.e. the directional (off-axis) and gratings-assisted (on-axis) couplings. In this article, we unify the investigation on the full-vectorial coupling principle for these two cases. The full-vectorial directional coupling is shown in Fig. 1. For example, we illustrate mode coupling from the fundamental mode $HE_{11}$ in general SMF to the first full-vectorial mode group in high-contrast-index ring-core fibers (RCFs) as the strongly guiding waveguides. The propagation constants must be equal for desired mode coupling, i.e., $\beta_1 = \beta_2$ as shown in Fig. 1(b), where $\beta = 2\pi n_{eff}/\lambda$ with $n_{eff}$ being effective refractive index and $\lambda$ being wavelength, and subscripts "1" and "2" denote two coupled modes. As for the full-vectorial directional coupling, the electric-field vector needs to be basically similar on the close side of two modes, shown as in Fig. 1(c), to maximize the coupling coefficient, which determines the polarization-dependent feature. It should be noticed that if the coupled full-vectorial modes degenerate into LP modes, the coupling case will degrade into higher-order LP mode coupling in FMFs [16,17,22].

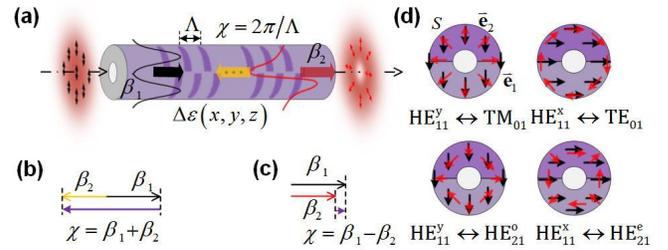

Fig. 2. The sketch and illustration of full-vectorial grating-assisted coupling. (a) Full-vectorial mode coupling from the fundamental mode $HE_{11}$ to the radially polarized mode $TM_{01}$ in RCF by optical gratings with both longitudinal and transverse non-uniform permittivity perturbation. The phase-matching condition of (b) reflection-type coupling and (c) transmission-type coupling. (d) The relationship between electric-field vector orientation of two coupled full-vectorial modes and transverse inhomogeneous permittivity perturbation.

As for on-axis full-vectorial coupling by optical gratings as shown in Fig. 2, the permittivity perturbation needs to be multidimensional function, and we assume that it can be separated as product of transverse and longitudinal modulation parts, i.e. $\Delta n(x,y,z) = \Delta n(x,y) \cdot \Delta n'(z)$. The longitudinal modulation $n'(z)$ is usually periodic function, i.e, $n'(z) \propto \cos(\chi z)$, giving the grating constant $\chi = 2\pi/\Lambda$ with $\Lambda$ being the modulation period to satisfy the phase-matching condition as shown in Fig. 2(b) and 2(c), corresponding to reflection and transmission coupling cases, respectively. The transverse inhomogeneous modulation $\Delta n(x,y)$ is responsible for breaking the orthogonality of two on-axis full-vectorial modes when making integral for Eq. (2). For example, the transverse refractive index modulation can be acoustically induced by $\Delta n(r,\phi) \propto r\cos\phi$ [23-25], and that for helical gratings can be written as $\Delta n(r,\phi) \propto \rho(r)\cos(\ell\phi)$, with $\rho(r)$ being the saturability of radial-dependent index modulation and $\ell$ is azimuthal order number [26,27]. In practice, the transverse index modulation can also be formed by tilted gratings [22,28], mechanical pressure [29] or femtosecond laser inscription [30]. It is noteworthy that this grating-assisted coupling also highly depends upon the electric-field vector orientation to maximize the coupling coefficient, which also reflects the polarization-dependent feature, as illustrated in Fig. 2(d).

### B. Coupled-mode equations

Based on the coupling coefficient analyzed above, for a specific polarization input, under faintly lifting mode degeneracy (i.e., the effective index difference between two full-vectorial modes is not large enough), the fundamental mode $HE_{11}$ in all coupling cases may be coupled to two full-vectorial modes, because of the no-zero coupling coefficients, as shown in Figs. 3 and 4. The thick violet arrow indicates the dominant coupling under full resonance condition, while the thin yellow ones indicate coupling crosstalk into other full-vectorial modes, because of approaching the phase matching condition. Note that the thin red arrows marked with cross indicate no coupling due to the zero coupling coefficient from Eq. (2). As for the three-mode coupling case, we give the general coupled-mode equations, as follows [9],

$$\frac{dA_1}{dz} = jK_{12}A_2 \cdot \exp\left[j(\beta_1 - \beta_2)z\right] + jK_{13}A_3 \cdot \exp\left[j(\beta_1 - \beta_3)z\right]$$

$$\frac{dA_2}{dz} = jK_{21}A_1 \cdot \exp\left[-j(\beta_1 - \beta_2)z\right]$$

$$\frac{dA_3}{dz} = jK_{31}A_1 \cdot \exp\left[-j(\beta_1 - \beta_3)z\right]$$

(3)

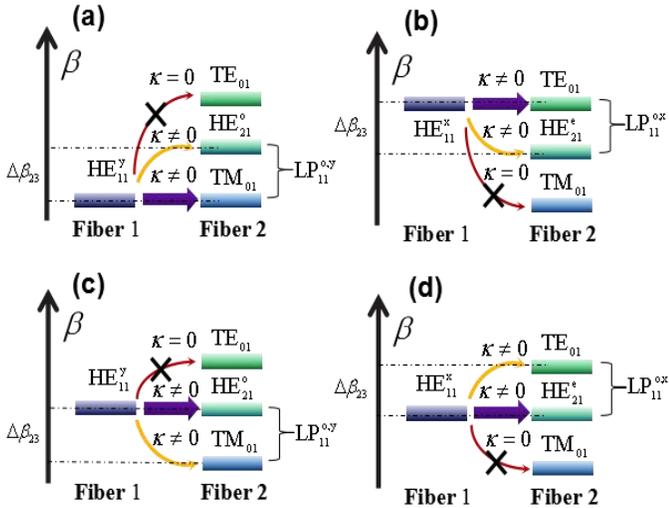

Fig. 3 Polarization-dependent directional coupling and coupling crosstalk under four full-vectorial mode coupling cases, corresponding to Fig. 1(c).

where $A_1$, $A_2$, and $A_3$ describe the complex electric field amplitudes of mode1 to mode3, respectively; $K_{12} = K_{21}$ indicates the reciprocal coupling coefficient between mode1 and mode2, and $K_{13} = K_{31}$ denotes that between mode1 and mode3, defined as Eq. (1). Specifically, as for the directional coupling, $K=\kappa$, being longitudinal-independent; while for the grating-assisted coupling, $K=\kappa \cdot \exp(\pm j\chi z)$ to give the phase resonant items of $\exp\left[\pm j(\beta_1 - \beta_k - \chi)z\right]$ in Eq. (3), where $k = 2$ and $3$. Here mode1 represents the fundamental mode $HE_{11}$, mode2 and mode3 correspond to the full resonant mode and crosstalk mode, respectively.

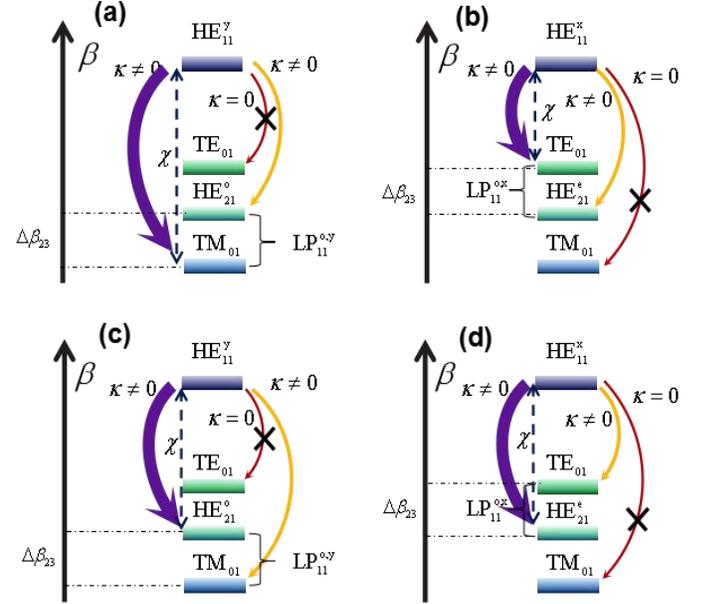

Fig. 4 Polarization-dependent grating-assisted coupling and coupling crosstalk under four full-vectorial coupling cases, corresponding to Fig. 2(d).

In the analysis model, we assume that the targeted full-vectorial coupling from mode1 to mode2 fully satisfies the phase matching condition. In other words, $\beta_1 = \beta_2$, and thus $\beta_1 - \beta_3 = \beta_2 - \beta_3 = \Delta\beta_{23}$ in the resonant items of Eq. (3) for the directional coupling case, while $\beta_1 - \beta_2 = \chi$, likewise, the resonant items $\beta_1 - \beta_3 - \chi = \Delta\beta_{23}$ for the grating-assisted coupling case. When redefining the electric field amplitudes of three coupled mode, $A_1 = T_1 \cdot \exp(j\Delta\beta_{23}z/2)$, $A_2 = T_2 \cdot \exp(j\Delta\beta_{23}z/2)$ and $A_3 = T_3 \cdot \exp(-j\Delta\beta_{23}z/2)$, the three mode-coupled equations in Eq. (3) can be transformed as,

$$\frac{dT_1}{dz} = -\frac{1}{2}j\Delta\beta_{23}T_1 + j\kappa_{12}T_2 + j\kappa_{13}T_3$$

$$\frac{dT_2}{dz} = -\frac{1}{2}j\Delta\beta_{23}T_2 + j\kappa_{12}T_1 \qquad (4)$$

$$\frac{dT_3}{dz} = \frac{1}{2}j\Delta\beta_{23}T_3 + j\kappa_{13}T_1$$

Notably, the mode degeneracy degree $\Delta\beta_{23}$ between mode2 and mode3 in Eq. (4) greatly influences the coupling crosstalk from mode1 to mode3, as shown in Figs. 3 and 4.

In the case of largely lifting mode degeneracy, i.e., $\Delta\beta_{23}$ is so large that the mode3 is completely detuned in Eq. (3) or (4). Accordingly, the three-mode coupling case would become the conventional two-mode coupling case. The normalized power evolution between mode1 and mode2 under the complete resonant condition can given by

$$P_1(z) = |A_1(z)|^2 / |A_1(0)|^2 = \cos^2(\kappa_{12} z), \quad (5)$$

$$P_2(z) = |A_2(z)|^2 / |A_1(0)|^2 = \sin^2(\kappa_{12} z). \quad (6)$$

Generally, in view of inadequately lifting mode degeneracy, $\Delta\beta_{23}$ is not large enough so as to produce nonnegligible coupling crosstalk from mode1 to mode3. It is difficult to give the analytical solutions of the three mode-coupled equations for Eq. (4). A numerical method can be available to solve them, as follows [34],

$$\frac{d}{dz}\mathbf{T}(z) = j \cdot \mathbf{M} \cdot \mathbf{T}(z), \quad (7)$$

where $\mathbf{T}(z) = [T_1 \ T_2 \ T_3]^T$ with superscript T meaning transpose, and the transmission matrix

$$\mathbf{M} = \begin{bmatrix} -\frac{1}{2}\Delta\beta_{23} & \kappa_{12} & \kappa_{13} \\ \kappa_{12} & -\frac{1}{2}\Delta\beta_{23} & 0 \\ \kappa_{13} & 0 & \frac{1}{2}\Delta\beta_{23} \end{bmatrix}. \quad (8)$$

Integrating Eq. (7) for the coupling length from 0 to $L$ with $L$ being the coupling length, it can be expressed as

$$\mathbf{T}(z) = \exp(j \cdot \mathbf{S}) \cdot \mathbf{T}(0), \quad (9)$$

with

$$\mathbf{S} = \int_0^L \mathbf{M} \cdot dz = L \cdot \mathbf{M}. \quad (10)$$

When diagonalizing the matrix $\mathbf{S}$ above, one can get a diagonal matrix $\mathbf{D}$ constituted by the eigen values of matrix $\mathbf{S}$, and the matrix $\mathbf{V}$ constituted by all the corresponding eigen vectors. The output matrix through the coupling region as the function of the input condition can be obtained as

$$\mathbf{T}(z) = \mathbf{V} \cdot \exp(j \cdot \mathbf{D}) \cdot \mathbf{V}^{-1} \cdot \mathbf{T}(0), \quad (11)$$

where $\mathbf{T}(0)$ is the initial input mode.

In the case of full mode degeneracy in the weakly guiding fibers, i.e., $\Delta\beta_{23} \simeq 0$, the normalized power of three coupled modes under the input from mode1 can be analytically solved as

$$P_1'(z) = |T_1(z)|^2 / |T_1(0)|^2 = \cos^2(\gamma z), \quad (12)$$

$$P_2'(z) = |T_2(z)|^2 / |T_1(0)|^2 = \frac{\kappa_{12}^2}{\gamma^2}\sin^2(\gamma z), \quad (13)$$

$$P_3'(z) = |T_3(z)|^2 / |T_1(0)|^2 = \frac{\kappa_{13}^2}{\gamma^2}\sin^2(\gamma z), \quad (14)$$

where $\gamma = \sqrt{\kappa_{12}^2 + \kappa_{13}^2}$, and $\kappa_{12} \simeq \kappa_{13}$. The mode2 and mode3 would be degenerated into LP modes with the total power of $P_2' + P_3' = \sin^2(\gamma z)$, coincident with the two mode coupling case that is given by Eq. (6). For both the directional and grating-assisted coupling in Figs. 1 and 2, respectively, we illustrate the full-vectorial mode component and meanwhile mode degeneracy into LP modes under the weakly guiding approximation, $\Delta\beta_{23} \simeq 0$, as shown in Fig. 5. Note that the case of coupling into $LP_{11}^{e,x}$ and $LP_{11}^{e,y}$ can be gotten by spatially rotating the fiber coupler structures with 90 degrees.

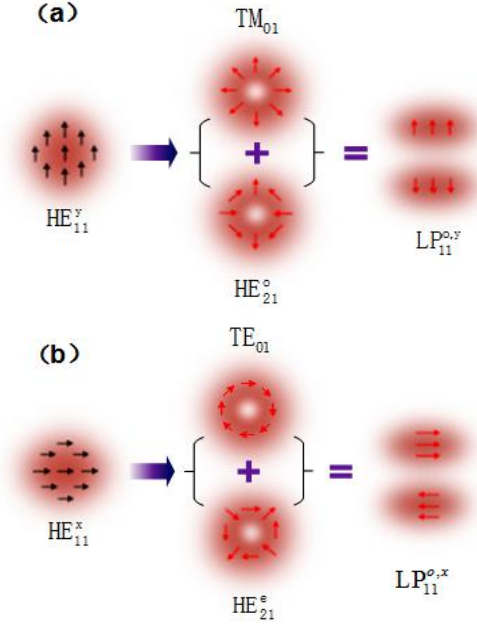

Fig. 5 Polarization-dependent full-vectorial mode component coupling and meanwhile mode degeneracy into (a) $LP_{11}^{o,y}$ and (b) $LP_{11}^{o,x}$ under weakly guiding approximation $\Delta\beta_{23} \simeq 0$.

III. SIMULATION RESULTS

In this part, we present the simulation results of full-vectorial mode coupling for both directional and grating-assisted coupling cases by means of the finite-difference time-domain (FDTD) solutions. In our design, the high-contrast-index RCF is adopted to lift the modal degeneracy [27,28]. The ring-core refractive index of RCF has a high step index of $\Delta \simeq 2.3\%$, relative to fiber cladding, corresponding to $n_2 = 1.478$. The cladding refractive index is 1.444, and the inner layer is air, i.e., $n_1 = 1$. The inner and outer radii of RCF are conditioned to $a_2 = 1.5a_1$. In this case, the effective index difference between any two full-vectorial modes approaches $1\times10^{-3}$, which is so large that there is scarcely any coupling crosstalk.

In the design, the full-vectorial directional coupler has the up and down core layout as shown in Fig. 1. The normal SMF has core radius $a_0 = 4.6$ μm and core refractive index $n_0 = 1.448$. The inner radius of RCFs is adjusted to satisfy the resonance condition as shown in Figs. 1(b). For full-vectorial coupling from the fundamental mode $HE_{11}$ in SMF to $TE_{01}$, $TM_{01}$ and $HE_{21}$ in RCF, the corresponding inner radii of RCF are $a_1 =$ 2554 nm, 2794 nm, and 2661 nm, respectively, at the wavelength of 1550 nm. The suitable core distances are set to make the coupling coefficient about 0.2 mm$^{-1}$ so that the full coupling length is about 8 mm. As for the full-vectorial grating-assisted coupler, it is designed as transverse non-uniform modulation with up and down dislocation as shown in Fig. 2. For the grating-assisted full-vectorial coupling, in our simulation, the RCF has the same parameters with the case of directional coupling for $TM_{01}$ mode. By the transmission-type grating-assisted coupling from $HE_{11}$ to $TE_{01}$, $TM_{01}$ and $HE_{21}$ modes in the same RCF, the corresponding grating periods are designed as $\Lambda$ = 4113.5 μm, 691.1 μm, and 1241.9 μm, respectively, based on the phase-matching condition shown in Fig. 2(c). A suitable modulation strength is set and the number of grating period is maintained about 45 so as to produce full coupling. Here we just simulate the transmission-type coupling case, the reflection-type coupling case is based on the similar coupling principle, apart from the design of grating period according to the reflection-type phase-matching condition as shown in Fig. 2(c).

Tab. 1. Numerical results about polarization-dependent directional excitation

| x/y-polarized $HE_{11}$ input | Resonance conditions and polarization-dependent coupling output | | |
| --- | --- | --- | --- |
| | $\beta_{HE_{11}} = \beta_{TE_{01}}$ | $\beta_{HE_{11}} = \beta_{TM_{01}}$ | $\beta_{HE_{11}} = \beta_{HE_{21}}$ |
| (image) | (image) | (image) | (image) |
| (image) | (image) | (image) | (image) |

Tab. 2. Numerical results about polarization-dependent excitation by gratings

| x/y-polarized $HE_{11}$ input | Phase-matching conditions and polarization-dependent coupling output | | |
| --- | --- | --- | --- |
| | $\beta_{HE_{11}} - \beta_{TE_{01}} = \chi$ | $\beta_{HE_{11}} - \beta_{TM_{01}} = \chi$ | $\beta_{HE_{11}} - \beta_{HE_{21}} = \chi$ |
| (image) | (image) | (image) | (image) |
| (image) | (image) | (image) | (image) |

The simulation results are exhibited in Tabs, 1 and 2, which clearly shows that the $TM_{01}$, $TE_{01}$ and $HE_{21}$ mode excitation highly depends upon the inputted polarization states. The $TE_{01}$ mode just can be excited by the x-polarized $HE_{11}$ input, whereas the $TM_{01}$ mode just can be excited by the y-polarized $HE_{11}$ input. Both the even and odd $HE_{21}$ modes can be excited by corresponding x and y polarized $HE_{11}$ modes, because the even and odd $HE_{21}$ modes have nearly the same effective refractive indices. Especially, it can be available to generate the circularly polarized vortex mode carrying OAM provided that the input mode has a circular polarization state [24,28,33]. The simulation results confirm that the full-vectorial mode coupling has the inherent polarization-dependence and -isolation features.

## IV. ANALYSES OF COUPLING CROSSTALK

In this section, we focus on analyses of coupling crosstalk for the full-vectorial mode coupling in view of the close effective refractive indices among the full-vectorial mode groups in optical fibers. We numerically solve the coupled-mode equations of three-mode coupling model by the methods of transmission matrix based on Eqs. (7)-(11). For convenience, approximately, we keep the coupling coefficients same, i.e., $\kappa_{12} = \kappa_{13} = \kappa$ in Eq. (8).

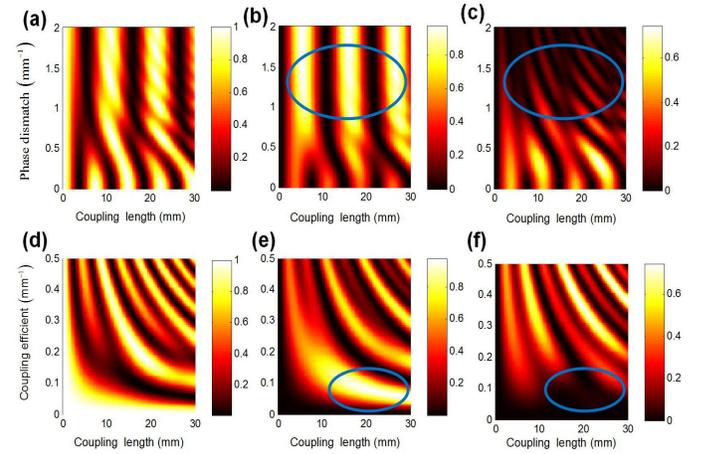

Fig. 6. Coupling evolution map versus coupling length under variable degrees of phase dismatch and coupling coefficients for three mode coupling. Power evolution versus phase dismatch degrees and coupling length for (a) mode1 as the input mode, (b) mode2 as targeted mode, and (c) mode3 as crosstalk mode under $\kappa = 0.3$ mm$^{-1}$. Power evolution versus coupling coefficients and coupling length for (d) mode1 as input mode, (e) mode2 as targeted mode, and (f) mode3 as crosstalk mode under $\Delta\beta_{23} = 0.4$ mm$^{-1}$

Firstly, we calculate the coupling efficiency of three modes under the variable degree of phase dismatch $\Delta\beta_{23}$ and different coupling length, as shown in Figs 6(a) to 6(c), corresponding to the input mode1, targeted coupled mode2, and mode3 as crosstalk mode, respectively. In this case, the coupling coefficient is fixed as a reasonable value of $\kappa = 0.3$ mm$^{-1}$. It shows that under the small $\Delta\beta_{23} \simeq 0$, the coupling crosstalk is so serious that nearly half of modal power is coupled into the crosstalk mode. As a result, the coupled mode2 and mode3 may be degenerated into LP modes [17,22], based on Eqs. (12)-(14). However, this coupling crosstalk can be sharply reduced by proverbially increasing the phase dismatch degree, as shown in the regions denoted with ellipses as contrast between Figs 6(b) and 6(c). When completely lifting the modal degeneracy, i.e. $\Delta\beta_{23} > 1$ mm$^{-1}$, the three-mode coupling case trends toward to the conventional two-mode coupling case with little coupling crosstalk, given by Eqs. (5) and (6). In addition, analogously, we give the coupling efficiency versus the variable coupling coefficient and coupling length when fixing the degree of phase dismatch at $\Delta\beta_{23} = 0.4$ mm$^{-1}$, as shown in Figs. 6(d) to 6(f). One can see that even in the case of small phase dismatch, the coupling crosstalk can also be efficiently reduced by decreasing the coupling coefficient and meanwhile increasing the coupling length.

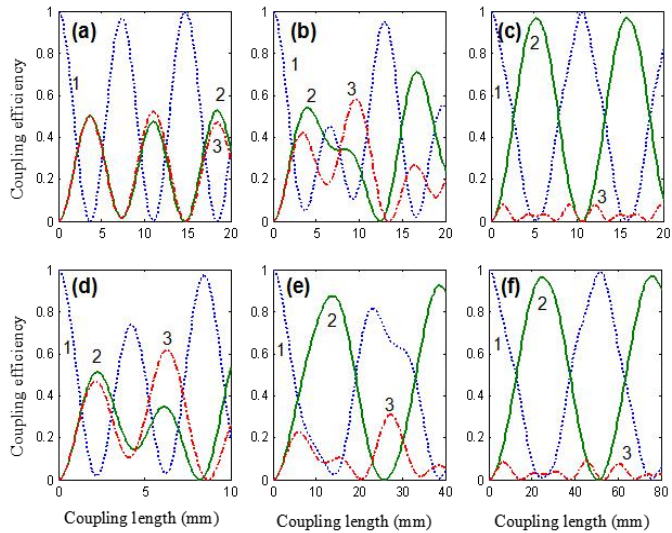

Fig. 7. Coupling efficiency from mode1 to mode2 and mode3 versus coupling length under different phase dismatch degrees and different coupling coefficients. (a) $\Delta\beta_{23} = 0.08$ mm$^{-1}$ and $\kappa = 0.3$ mm$^{-1}$, (b) $\Delta\beta_{23} = 0.4$ mm$^{-1}$ and $\kappa = 0.3$ mm$^{-1}$, (c) $\Delta\beta_{23} = 2.0$ mm$^{-1}$ and $\kappa = 0.3$ mm$^{-1}$, (d) $\Delta\beta_{23} = 0.4$ mm$^{-1}$ and $\kappa = 0.5$ mm$^{-1}$, (e) $\Delta\beta_{23} = 0.4$ mm$^{-1}$ and $\kappa = 0.125$ mm$^{-1}$, and (f) $\Delta\beta_{23} = 0.4$ mm$^{-1}$ and $\kappa = 0.0625$ mm$^{-1}$.

To better exhibit the results of coupling crosstalk reduction, we further plot the power evolution curves of three mode coupling, when fixing $\kappa = 0.3$ mm$^{-1}$, and increasing $\Delta\beta_{23}$ to 0.08 mm$^{-1}$, 0.4 mm$^{-1}$, and 2.0 mm$^{-1}$, corresponding to Figs. 7(a) to 7(c), respectively; and fixing $\Delta\beta_{23} = 0.4$ mm$^{-1}$, and decreasing $\kappa$ to 0.5 mm$^{-1}$, 0.125 mm$^{-1}$, and 0.0625 mm$^{-1}$, corresponding to Figs. 7(d) to 7(f), respectively. From Fig. 7, it clearly shows that the coupling efficiency from mode1 to mode2 can be sharply improved with reducing the coupling crosstalk to mode3 by not only enhancing the phase dismatch degree $\Delta\beta_{23}$ but also decreasing the coupling coefficient. In practice, the phase dismatch degree can be enlarged by increasing the refractive index contrast between fiber core and cladding to fully lift the modal degeneracy. The coupling coefficient for the full-vectorial directional coupling can be weakened by enlarging the distance between two fiber cores to decrease the overlap region of two coupled modes, while for the full-vectorial grating-assisted coupling, it can be weakened by adopting the weak permittivity perturbation when fabricating fiber gratings.

## V. Conclusion

We have uncovered the mode coupling principle of fiber-guided full-vectorial modes. It is characterized by the inherent polarization-dependence and polarization-isolation coupling features, which is different from the conventional linear polarization (LP). Furthermore, the coupling crosstalk of full-vectorial mode coupling is analyzed in detail in view of the modal degeneracy degree based on the three-mode coupling model. We firstly find that the coupling crosstalk can be efficiently reduced by means of not only increasing the degree of lifting modal degeneracy, but also decreasing the coupling coefficient and meanwhile increasing coupling length. Our findings are generally suitable for all full-vectorial mode coupling cases in all kinds of optical fibers. It is expected that the revealed full-vectorial coupling mechanism might find a significant guidance on exploiting the full-vectorial modes in the realms of vectorial-mode-based optical signal processing, communications and sensing systems.

## Acknowledgment

This work was supported by the National Natural Science Foundation of China (NSFC) under grants 11574001, 11274131 and 61222502, the Program for New Century Excellent Talents in University (NCET-11-0182), the Wuhan Science and Technology Plan Project under grant 2014070404010201, and the seed project of Wuhan National Laboratory for Optoelectronics (WNLO).

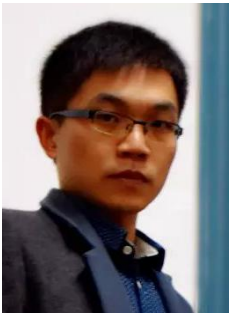
Liang Fang received the M.S. degree in Optical Engineering from University of Shanghai for Science and Technology, Shanghai, China, in 2015, where he focused his research on optical fiber gratings and mode-division multiplexing. He is currently pursuing his Ph.D. Degree in Optical Engineering at the Wuhan National Laboratory for Optoelectronics, Huazhong University of Science and Technology, Wuhan, China, where his research interest includes optical waveguides and mode coupling, angular momentum and Doppler effect of structured light, and photonics spin Hall effect in micro/nano waveguide. He has published more than 10 journal papers on Laser & Photonics Reviews, Physical Review A, Optics Express, Optics Letters, etc.

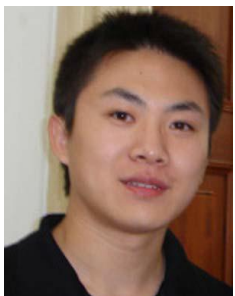
Jian Wang received the Ph.D. degree in physical electronics from the Wuhan National Laboratory for Optoelectronics, Huazhong University of Science and Technology, Wuhan, China, in 2008. He worked as a Postdoctoral Research Associate in the Ming Hsieh Department of Electrical Engineering, University of Southern California, USA, from 2009 to 2011. He is currently a professor at the Wuhan National Laboratory for Optoelectronics, Huazhong University of Science and Technology, Wuhan, China. He has devoted his research efforts to innovations in photonic integrated devices and frontiers of optical communications and optical signal processing. He has more than 300 publications, including 3 book chapters, 2 special issues, 3 review articles, 5 invited papers, 42 tutorial/keynote/invited talks (invited talk at OFC2014, tutorial talk at OFC2016), 8 postdeadline papers, and more than 100 journal papers published on Science, Nature Photonics, Scientific Reports, Applied Physics Letters, Optics Express, Optics Letters, etc.